\documentclass[traditabstract, rnote]{aa}
\usepackage[utf8]{inputenc}
\usepackage{setspace}
\usepackage{amsmath}
\usepackage{amssymb}
\usepackage{microtype}
\usepackage{natbib}
\usepackage{graphicx}
\usepackage{txfonts}
\usepackage{color}
\usepackage{multirow}
\usepackage{xspace}
\bibpunct{(}{)}{;}{a}{}{,} % to follow the A&A style
\newcommand{\es}{\,erg\,s$^{-1}$\xspace}

\title{Searching for coherent pulsations in
ultraluminous X-ray sources}
\author{V.\,Doroshenko\inst{1}, A.\,Santangelo\inst{1}, L.\,Ducci\inst{1,2}}
\institute{Institut für Astronomie und Astrophysik, Sand 1, 72076 Tübingen, Germany, \email{doroshv@astro.uni-tuebingen.de}\and
ISDC, Department of Astronomy,  Universit\'e de Gen\`eve, chemin d'\'Ecogia, 16, CH-1290 Versoix, Switzerland}

\begin{document}

\bibliographystyle{aa}

\abstract{Luminosities of ultraluminous X-ray sources (ULXs) are uncomfortably large if compared to the Eddington limit for isotropic accretion onto stellar-mass object. Most
often either supercritical accretion onto stellar mass black hole or accretion onto intermediate mass black holes is invoked the high luminosities of ULXs. However, the recent
discovery of coherent pulsations from M82 ULX with \emph{NuSTAR} showed that another scenario implying accretion onto a magnetized neutron star is possible for ULXs. Motivated by
this discovery, we re-visited the available \emph{XMM-Newton} archival observations of several bright ULXs with a targeted search for pulsations to check whether accreting neutron
stars might power other ULXs as well. We have found no evidence for significant coherent pulsations in any of the sources including
the M82 ULX. We provide upper limits for the amplitude of possibly undetected pulsed signal for the sources in the sample.}

% TODO ADD Keywords
\keywords{X-rays: binaries, X-rays: galaxies, Stars: neutron, Stars: black holes}
\authorrunning{V. Doroshenko et al.}
\maketitle

\section{Introduction} Ultraluminous X-ray sources (ULX) are vaguely defined as
non-nuclear extragalactic sources with luminosities exceeding
$\sim10^{39}$\es, i.e. the Eddington limit for isotropic accretion on
stellar-mass black hole \citep{Feng11}. These sources are considered, therefore,
best candidates to host intermediate mass black holes (IMBHs,
\citealp{Colbert99}). Alternatively, super-eddington accretion onto stellar
mass black holes like in the case of the well known source SS~433 \citep{King01,
Fabrika04}. Emission from young rotation-powered pulsars
\citep{Medvedev13} has been suggested as a possible mechanism powering ULXs.

Recently, the discovery of coherent X-ray pulsations from ULX M82 X-2 (or X-1) \citep{Bachetti14}, with period of $\sim1.37$\,s and the associated orbital motion, provided a new
explanation for the very bright emission of at least some of these objects. Based on the X-ray timing \cite{Bachetti14} have derived a mass function of $f=2.1 M_\odot$. The
accretion would then proceed from a low mass companion via Roche-lobe overflow, likely the only mechanism to feed enough matter to explain the observed luminosity. This
implies a strong spin-up torque imposed onto the neutron star in agreement with the observed short spin-up timescale ($P/\dot{P}\sim300\,{\rm yr}$) and spin period of the pulsar.
In principle, other ULXs powered by an accreting neutron star could exist and are expected to have spin periods of the order $\lesssim10$\,s as well.

Currently there is a lack of detailed studies of the variability in ULXs at
short timescales as illustrated by the surprising discovery of the pulsations
in M82 ULX, which is a rather well studied system. Motivated by this,
we performed a targeted search for coherent pulsations with periods in range
0.15-15\,s in archival \emph{XMM-Newton} data to understand whether other ULXs might
host accreting pulsars. In the present note, we report results of this systematic search.

\section{Source sample and available data} The number of ULXs and
candidates is steadily increasing since the launch of the \emph{Einstein}
satellite. Many of them are, however, not sufficiently bright for detailed
timing analysis. As an initial step, we limited our analysis to a sample of
fifteen bright ULXs observed with \emph{XMM-Newton}. This sample was used by
\cite{Heil09} to characterize ULX variability in a broad frequency range and no
significant periodicities were reported by these authors. However,
\cite{Heil09} did not search for coherent pulsations.

We searched for pulsation in archival \emph{XMM-Newton} data of the EPIC PN camera as it is one of the few instruments with adequate timing resolution to investigate variability up to
sub-second timescales. Our analysis is based on a larger dataset compared to the one considered by \cite{Heil09} since many additional observations have become available since
their publication. Most of the observations were performed in full-frame read-out mode with time resolution of $\sim0.07$\,s. The list of the sources in our sample, and the
summary of available observations are presented in Table~\ref{tab:summary}.

\section{Data analysis and results} Low-level data reduction was carried
out using the \emph{XMM} SAS 13.5 package, current calibration files and standard
filtering criteria\footnote{http://xmm.esac.esa.int/sas/current/doc}. 
Source photons with energies in 0.3-10\,keV range were then extracted using the
source-centered circles with radius of 20\arcsec in all cases except M82, where
we used a radius of 40\arcsec centered in between the two nearby ULXs, X-1 and
X-2, where the pulsations could potentially originate from \citep{Bachetti14}.
The arrival times of individual photons were then corrected to solar barycenter
using the \emph{barycen} task.

To search for pulsations, we used the H-test \citep{Jager89} applied to
unbinned source events. Taking into account that in ULX M82 the pulsed fraction
showed large variations with time (suggesting the possible transient nature of the
pulsations), we analyzed individual observations separately. 
Results are
summarized in Table~\ref{tab:summary}.

We have found no evidence for significant periodic signals in any of the
sources. Therefore, we followed the approach suggested by \cite{Brazier94} to
derive the upper limits on amplitude of potentially present but undetected
periodic signal $f_{pulsed}$. For each source and observation we calculated the
upper-limit on the amplitude of a sinusoidal signal with period in the range
0.15-15 s at 3$\sigma$ confidence level (for detail see Eq.~3 and accompaining text in
\citealp{Brazier94}). We report in Table~\ref{tab:summary} the highest
upper-limit among all the observations available for a given source.
% In Table~\ref{tab:summary}
% for each source we list the (strongest among all observations for given source)
% upper limits on amplitude of sinusoidal signal with period in range 0.15-15\,s
% at $3\sigma$ confidence level (see Eq.~3 and accompaining text in
% \cite{Brazier94} for detail).
Note that for ULX in M82 our limit is below the
lowest value reported by \cite{Bachetti14}.

\begin{figure}[ht]
	\centering
		\includegraphics[width=0.5\textwidth]{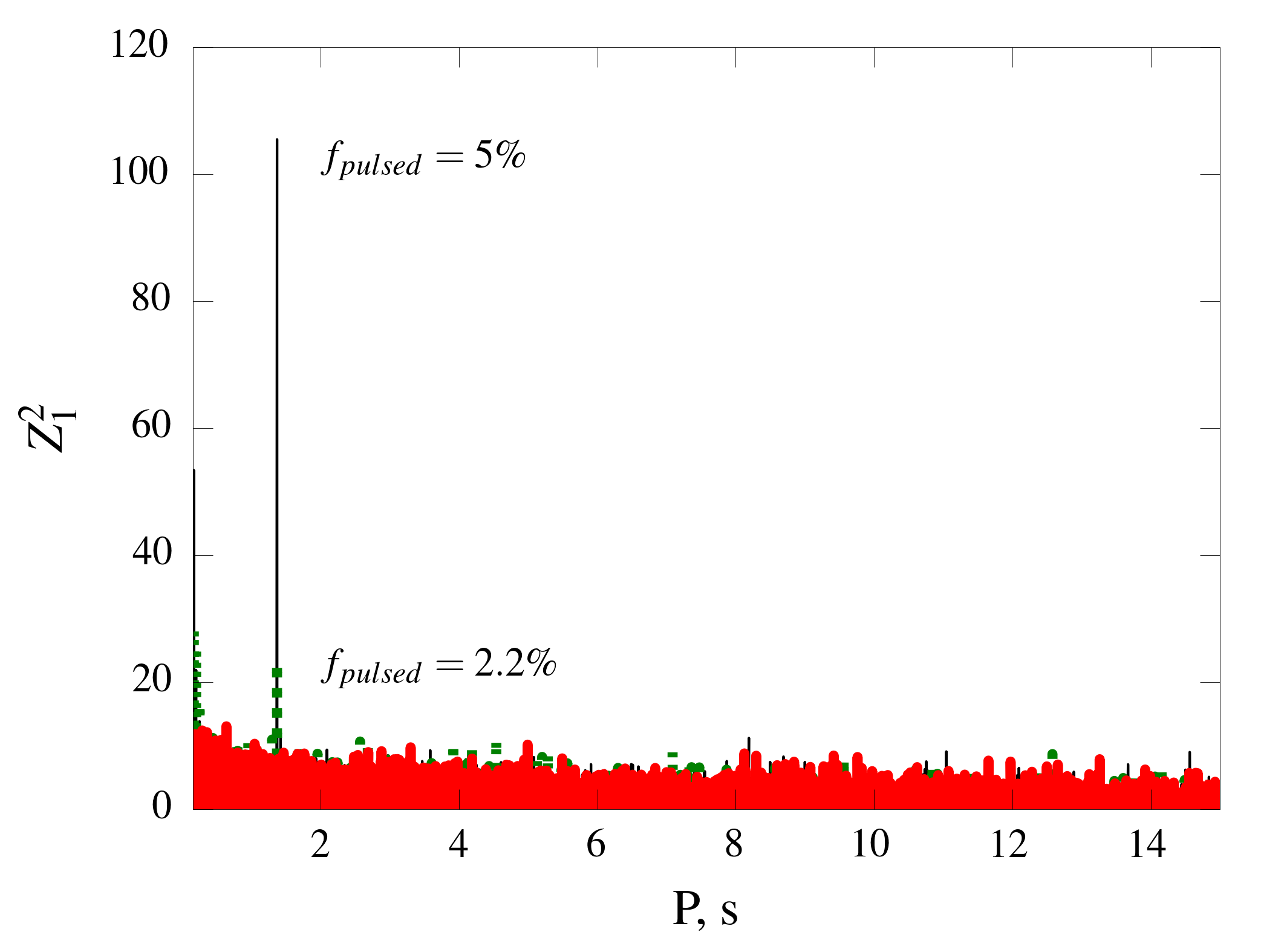}
	\caption{A periodogram for the longest observation of the pulsating in ULX in M82 (thick red
line, obsid 0206080101, exposure of 46\,ks), and periodograms for simulated
sinusoidal signal with pulsed fractions of 5\% (thin black line, only visible
at the peak) and 2.2\% (dotted green line) with same exposure, observation
duration and total number of photons as observed.}
	\label{fig:m82}
\end{figure}

A periodogram for the longest \emph{XMM} observation of the pulsating ULX in M82 
and periodograms for simulated sinusoidal signals with period of 1.37\,s and
pulsed fractions of 2.2\%, and 5\% is presented in Fig.~\ref{fig:m82}. Simulated
signals correspond to the derived upper limit for this observation, and to the
lowest amplitude reported for \emph{NuSTAR} in 0.3-10\,keV energy range. Note,
that in both cases pulsations are clearly detected (although with less than
3$\sigma$ significance for $f_{pulse}=2.2\%$). Surprisingly, no signal is evident in the
\emph{XMM} data. 

	% \begin{center}
\begin{table}[!hb]
	\caption{List of sources included in the analysis. Upper limits on pulsed fraction represent strongest (among all observations) limits at $3\sigma$ confidence level.}
	\begin{tabular}{llllll}                                                                                      
         Source & Exp.(ks)/obs. &  $f_{pulsed}$ limit, \%\\
\hline
    M82 X-1/X-2 &        143/11 &                   2.2\\
     NGC 55 ULX &         130/3 &                   4.4\\
  NGC 253 PSX-2 &          69/7 &                   9.8\\
   NGC 1313 X-1 &        315/18 &                   4.4\\
   NGC 2403 X-1 &          66/4 &                   8.8\\
Holmberg II X-1 &          76/7 &                   3.7\\
Holmberg IX X-1 &        195/13 &                   4.6\\
   NGC 3628 X-1 &          40/1 &                  17.1\\
   NGC 4395 X-1 &          91/3 &                   8.6\\
   NGC 4559 X-1 &          31/1 &                  11.7\\
   NGC 4861 ULX &          19/3 &                  30.7\\
   NGC 4945 X-2 &          39/2 &                  25.7\\
   NGC 5204 X-1 &          82/8 &                   8.1\\
        M83 ULX &          31/2 &                  30.5\\
   NGC 5408 X-1 &        473/10 &                   3.6\\
\hline
	\end{tabular}
	\label{tab:summary}
\end{table}
	% \end{center}

\section{Conclusions} Inspired by the recent discovery of the pulsations from the
ULX in M82 with \emph{NuStar}, we revisited the available archival \emph{XMM-Newton}
observations of several bright ULXs in order to systematically search for
pulsations whoose detection escaped previous investigations (as it was the case for
ULX in M82). We found no significant pulsed signal in range of periods from 0.15
to 15\,s in any of the considered sources including the ULX in M82. We provide,
therefore, upper limits for pulsed fraction of potentially non-detected
pulsations. Note, that in many cases these limits are rather weak due to
limited statistics and could be significantly improved with additional observations.

For ULX M82 our upper limit turns to be a factor of two lower than the lowest
value $f_{pulsed}\sim5$\% reported by \cite{Bachetti14}, and therefore, we
exclude pulsations with amplitude similar to the one observed by
\emph{NuSTAR} in the \emph{XMM-Newton} data. The amplitude of the pulsations in
this source, however, has been reported to vary with time, hence it can not be
excluded that at the time of the \emph{XMM-Newton} observation it remained intrinsically low. This would imply that pulsations in ULXs powered by
accreting neutron stars might be transient and highlights the importance of
regular monitoring of ULXs, particularly at higher energies where pulsed
fraction is expected to be larger. Still, an independent confirmation of
pulsations in ULX in M82 would be indispensable.

\begin{acknowledgements}
authors thank the Deutsches Zentrums für Luft- und Raumfahrt (DLR) and Deutsche
Forschungsgemeinschaft (DFG) for financial support (grants DLR~50~OR~0702, FKZ
50 OG 1301, SA2131/1-1).
\end{acknowledgements}

\bibliography{biblio}

\begin{thebibliography}{9}
\expandafter\ifx\csname natexlab\endcsname\relax\def\natexlab#1{#1}\fi

\bibitem[{Bachetti {et~al.}(2014)Bachetti, Harrison, Walton, Grefenstette,
  Chakrabarty, Furst, Barret, Beloborodov, Boggs, Christensen, Craig, Fabian,
  Hailey, Hornschemeier, Kaspi, Kulkarni, Maccarone, Miller, Rana, Stern,
  Tendulkar, Tomsick, Webb, \& Zhang}]{Bachetti14}
Bachetti, M., Harrison, F.~A., Walton, D.~J., {et~al.} 2014, Nature, 514, 202

\bibitem[{{Brazier}(1994)}]{Brazier94}
{Brazier}, K.~T.~S. 1994, \mnras, 268, 709

\bibitem[{{Colbert} \& {Mushotzky}(1999)}]{Colbert99}
{Colbert}, E.~J.~M. \& {Mushotzky}, R.~F. 1999, \apj, 519, 89

\bibitem[{{de Jager} {et~al.}(1989){de Jager}, {Raubenheimer}, \&
  {Swanepoel}}]{Jager89}
{de Jager}, O.~C., {Raubenheimer}, B.~C., \& {Swanepoel}, J.~W.~H. 1989, \aap,
  221, 180

\bibitem[{{Fabrika}(2004)}]{Fabrika04}
{Fabrika}, S. 2004, Astrophysics and Space Physics Reviews, 12, 1

\bibitem[{{Feng} \& {Soria}(2011)}]{Feng11}
{Feng}, H. \& {Soria}, R. 2011, \nar, 55, 166

\bibitem[{{Heil} {et~al.}(2009){Heil}, {Vaughan}, \& {Roberts}}]{Heil09}
{Heil}, L.~M., {Vaughan}, S., \& {Roberts}, T.~P. 2009, \mnras, 397, 1061

\bibitem[{{King} {et~al.}(2001){King}, {Davies}, {Ward}, {Fabbiano}, \&
  {Elvis}}]{King01}
{King}, A.~R., {Davies}, M.~B., {Ward}, M.~J., {Fabbiano}, G., \& {Elvis}, M.
  2001, \apjl, 552, L109

\bibitem[{{Medvedev} \& {Poutanen}(2013)}]{Medvedev13}
{Medvedev}, A.~S. \& {Poutanen}, J. 2013, \mnras, 431, 2690

\end{thebibliography}
\end{document}